\documentclass[prd,aps,floats,twocolumn,nofootinbib]{revtex4-1}
\usepackage{slashed}
\usepackage{mathtools}
\usepackage{amsfonts}
\usepackage{amssymb}
\usepackage{epsfig}
\usepackage{bm}

\begin{document}

\newcommand{\m}[1]{\mathcal{#1}}
\newcommand{\nn}{\nonumber}
\newcommand{\ph}{\phantom}
\newcommand{\eps}{\epsilon}
\newcommand{\be}{\begin{equation}}
\newcommand{\ee}{\end{equation}}
\newcommand{\bea}{\begin{eqnarray}}
\newcommand{\eea}{\end{eqnarray}}
\newtheorem{conj}{Conjecture}
\newcommand{\plk}{\mathfrak{h}}


\title{Unimodular Proca Theory: Breaking the $U(1)$ gauge symmetry of unimodular gravity via a mass term}
\date{16/05/2023}

\author{Raymond Isichei}
\author{Jo\~{a}o Magueijo}
\email{ri221@imperial.ac.uk}
\email{j.magueijo@imperial.ac.uk}
\affiliation{Theoretical Physics Group, The Blackett Laboratory, Imperial College, Prince Consort Rd., London, SW7 2BZ, United Kingdom}

\begin{abstract}
We study the Hamiltonian structure of unimodular-like theories, where the cosmological constant (or other supposed constants of nature) are demoted from fixed parameters to classical constants of motion. No new local degrees of freedom are present as a result of a $U(1)$ gauge invariance of the theory. Hamiltonian analysis of the action reveals that the only possible gauge fixing that can be enforced is setting the spatial components of the four-volume time vector ${\cal T}^{i}\approx0$. As a consequence of this, the gauge-fixed unimodular path integral is equivalent to the minisuperspace unimodular path integral. However, should we break the $U(1)$ gauge invariance, two things happen: a massless propagating degree of freedom appears, and the (gauge-invariant) zero-mode receives modified dynamics. The implications are investigated, with the phenomenology depending crucially on the target ``constant''.
\end{abstract}

\maketitle

\section{Introduction}
Ever since General Relativity was proposed, a question has been hanging over the theory, as posed by Einstein himself~\cite{Einstein-unimod}: should diffeomorphisms be restricted to volume preserving ones? If so, this assumption leads to the so-called ``unimodular'' theory of gravity, and this was one of the first attempts at grand-unification. Later it was found that 
unimodular gravity demotes the cosmological constant from a fixed pre-given parameter to a constant of motion~\cite{unimod1,unimod,UnimodLee1,alan,daughton,sorkin1,sorkin2,vikman,vikman1,vikmanaxion},  leading to considerable ambiguity~\cite{unimod1,unimod,UnimodLee1,unimod3} over whether this resolves, ameliorates or simply does nothing to solve the cosmological constant problem~\cite{weinberg,padilla-review}.

Possibly the cleanest formulation of unimodular gravity is due to Henneaux and Teitelboim (HT)~\cite{unimod}, who eschew the restriction of fixed volume for the diffeomorphisms in favour of a Lagrange multiplier density ${\cal T}^\mu$ enforcing the on-shell constancy of $\Lambda$ (preserving full diffeomorphism invariance). This density can be used to build a physical definition of time, which on-shell becomes ``4-volume time''~\cite{Bombelli,sorkin1,sorkin2,UnimodLee2}. The HT procedure can be used as a blueprint for turning other parameters into constants of motion~\cite{JoaoLetter,JoaoPaper}, such as the Planck mass or the gravitational coupling, with alternative conjugate time variables, such as the Ricci or the matter times, respectively.

Unimodular gravity does not introduce new local degrees of freedom, even though it does introduce a global one. 
The reason for this is a $U(1)$ gauge-invariance of the theory (first noted by~\cite{unimod}, and with several reformulations which we review in Section~\ref{symmetry}). As the Hamiltonian analysis in Section~\ref{HamHT} will show, this induces constraints which always work to subtract from the phase space any new local degrees of freedom, whatever formulation one chooses. This is also evident in the path integral formulation (Section~\ref{pathint}). 

But if we break the $U(1)$ gauge-invariance, for example, via a Proca mass term (as done for electromagnetism in Maxwell-Proca theory), this releases a new local particle, as we show in this paper. We first do this in Section~\ref{Proca-Solutions} by explicitly evaluating the equations of motion and their solutions. These can be split into a revised zero-mode equation (producing a time-varying 
Lambda, instead of the ususal on-shell constant, and a corrected conjugate relational time) together with a propagating massless mode. The associated phenomenology is not ideal, at least for the simplest models (as we discuss in Section~\ref{discussion}), but the purpose of this paper is to illustrate the concept, rather than to fine-tune the phenomenology. Also the observational implications do depend on the target constant. 

In this paper we shall use signature $-+++$ (important for defining the sign of the Proca term) and use units such that $\hbar=c=1$ (relevant for evaluating the mass dimensions of the Proca coupling). 


\section{The HT formulation of unimodular theory}
In the HT formulation of unimodular theory
full diffeomorphism invariance is preserved, but one adds to the base action $S_0$  an additional term:
\be\label{Utrick}
S_0\rightarrow S=S_0+ S_U\equiv 
S_0 - \int d^4 x \, \Lambda(\partial_{\mu}{\cal T}^\mu).
\ee 
Here  ${\cal T}^\mu$ is a  density, so that the added term is diffeomorphism invariant without using the metric or the connection.  Since these do not appear in the new term, the Einstein equations (and other
field equations) are left unchanged. 
In standard HT, given that $S_0$ does not depend on ${\cal T}^\mu$, one gets 
on-shell constancy of $\Lambda$. 
Note the gauge  symmetry~\cite{unimod}:
\begin{equation}\label{gauge}
    {\cal T}^\mu\rightarrow {\cal T}^\mu+\epsilon^\mu\quad {\rm with}\quad \partial_\mu\epsilon^\mu =0,
\end{equation}
rendering local degrees of freedom in the theory pure gauge modes. 
However, the zero-mode of ${\cal T}^0$:
\begin{equation}\label{zeroT}
    T(\Sigma)\equiv \int_{\Sigma}  d^3 x\, {\cal T}^0
\end{equation}
is gauge-invariant. It provides a physical definition of time, canonically dual to $\Lambda$. On-shell $\cal{T}$ is proportional to 
the 4-volume to the past, or unimodular time (indeed it is equal to it if we replace $\Lambda$ by $\rho_\Lambda$ in the above).

More generally, we may select a set of $D$ constants $\bm\alpha$ and take: 
\begin{equation}\label{genunimod}
 S_0\rightarrow S=S_0 - \int d^4 x\, {\bm \alpha} \cdot  \partial_{\mu}{ \cal T}_{\bm \alpha}^\mu
\end{equation}
where the dot denotes the Euclidean inner product in $D$ dimensional space. As with unimodular theory, the zero-modes of the zero components of the density $\mathbf{ \cal T}_{\bm \alpha}^\mu$ provide definitions of time $\bm T_{\bm\alpha}$, dual to the on-shell constants $\bm\alpha$. We will also occasionally integrate the unimodular action by parts and use:
\begin{equation}
    S_0\rightarrow S=S_0 + \int d^4 x\, (\partial_{\mu} {\bm \alpha}) \cdot  { \cal T}_{\bm \alpha}^\mu
\end{equation}
(but see the implications this has for the path integral in~\cite{PathInt}). 

For definiteness we will make our arguments with $\Lambda$, but at the end of the paper will contemplate more general situations. Where required we will also choose:
\begin{equation}
    S_0=\frac{M_P^2}{2}\int d^4 x\, {\sqrt{-g}} (R-2\Lambda)
\end{equation}
for definiteness but any alternative  theory of gravity could be inserted into our calculations. Here $M_P^2=1/(8\pi G)$ is the reduced Planck mass.

\section{The underlying symmetry}\label{symmetry}
The underlying symmetry of unimodular-like theories is a non-standard representation of the $U(1)$ symmetry, which has appeared in the literature in various contexts. In spite of the superficial differences, its various guises are closely related.

By dualizing $\cal T^\mu$ we obtain a 3-form density, $\mathbf{T}$, in terms of which the HT
action can be written as:
\begin{equation}
    S_U=\int \Lambda d{\mathbf{ T}}=\int \Lambda \mathbf{F}.
\end{equation}
The gauge transformation (\ref{gauge}) has the dual formulation $\mathbf{T}\rightarrow  \mathbf{T}+ \mathbf{E}$ where $d\mathbf{E}=0$ (so that\footnote{Recall that any p-form can be written as the sum of a closed p-form and the dual of a closed $D-p$-form.} $\mathbf{E}=d\mathbf{\Omega}_2$ for a 2-form $\mathbf{\Omega}_2$), resulting in:
\begin{eqnarray}
     \mathbf{T}&\rightarrow &   \mathbf{T}+ d \mathbf{\Omega}_2.
\end{eqnarray}
This is just the usual $U(1)$ gauge transformation represented by a 3-form gauge-field, with the usual scalar function replaced by a 2-form $\mathbf{\Omega_2}$. It is a 3 dimensional representation  because we can shift $\mathbf{\Omega_2}\rightarrow \mathbf{\Omega_2} + d \mathbf{\Omega_1}$, where in turn we can shift $\mathbf{\Omega_1}\rightarrow \mathbf{\Omega_1} + d \mathbf{\Omega_0}$. So of the 4 degrees of freedom of $\mathbf{\Omega_1}$ only 3 are relevant, and so of the 6 degrees of freedom of $\mathbf{\Omega_2}$ only 3 are relevant.

Relation with Hawking's electromagnetic 3-form~\cite{Hawking3form}
(and related supergravity constructions~\cite{Aurilia}) is evident. Ditto with p-form Electromagnetism~\cite{HTEMpform}, which amounts to an equivalent construction but with membranes as charged sources. But superficially the unimodular action seems different. The two differences are that Hawking's action contains the metric, whereas the unimodular one does not; and that the Hawking's action is quadratic in F (it is an electromagnetic action; cf. $p$-form EM as in~\cite{HTEMpform}), whereas the unimodular action is linear in $F$.

Regarding the first difference we note that Ref.~\cite{Bufalo} has also elected to define unimodular theory with the metric included:
\begin{eqnarray}\label{unimodVec}
    S_U=\int d^4 x\, \sqrt{-g} T^\mu \partial _\mu \Lambda
\end{eqnarray}
where $T^\mu$ is a proper vector (or a density with weight zero). This is equivalent to defining $T^\mu = \sqrt{-g}{\cal T}^\mu$ and
does not have an effect on the classical theory without Proca symmetry breaking terms. Indeed:
\begin{equation}
   S_U= -\int d^4 x\Lambda \partial_\mu {\cal T}^\mu = -\int d^4 x\sqrt{-g} \Lambda \nabla _\mu  T^\mu 
\end{equation}
with the equivalent equations of motion 
\bea
\nabla_\mu T^\mu&=& -M_P^2\label{UEOM1}\\
\partial_\mu \Lambda&=&0.\label{UEOM2}
\eea
The contribution to the energy-momentum tensor of the new term in the action is proportional to $\partial_\mu \Lambda$ and so vanishes on-shell. So although in this formulation the unimodular action contains the metric, the new term does not contribute to the Einstein equations. We will return to this formulation presently. 

Regarding the second difference, we could regard unimodular and p-form electromagnetism as ``$F(R)$'' (in analogy with the Einstein-Hilbert action) versions of each other, since one is quadratic in $F$, the other linear. This is not very useful, since $F$ is essentially a scalar (it is dual to a scalar), so at least in the most basic case all these theories are just field redefinitions of each other. This is not true in more complicated models, or if one breaks the gauge symmetry.

\section{The Hamiltonian structure of HT unimodular theory}\label{HamHT}
As was traditional at the time, the HT action was reconstructed from the Hamiltonian and the action's presumed 3+1 split, but we can work in the opposite direction (as was done before in~\cite{UnimodLee1}). In doing so, many different treatments are possible, all leading to the same result: that no local degrees of freedom arise. The multitude of possible treatments is due to the freedom to integrate the unimodular extension by parts. The resulting boundary terms are constants which do not affect the Hamiltonian analysis. 

\subsection{Method 1}

For example, if the derivatives are on $\Lambda$ then we may consider the integral
\be
S_U=
\int d^4 x \, (\partial_\mu \Lambda ){\cal T}^\mu=\int dt\, d^3 x[\dot \Lambda {\cal T}^0+ (\partial_i\Lambda) {\cal T}^ i ]
\ee
and immediately state that we have 2 new phase space variables ($\Lambda$ and  ${\cal T}^0$), and look at ${\cal T}^i$ as Lagrange multipliers enforcing one new primary constraint
\begin{equation}\label{const1}
    \partial_i\Lambda\approx 0.
\end{equation}
This does not generate secondary constraints, and it is first class because it commutes with everything else.
Hence the counting of local degrees of freedom goes as:
\begin{equation}
    D=\frac{2-2\times1}{2}=0.
\end{equation}
However, there are more complicated alternatives which, whilst exactly equivalent to this simple argument for straight unimodular theory, will be helpful to clarify matters once we break its gauge invariance.

\subsection{Method 2}\label{method2}
Although a momentum conjugate to ${\cal T}^{i}$ does not appear in the 3+1 split unimodular action, this can be inserted in the split action with  Lagrange multipliers $\lambda^{i}$ forcing it to vanish on-shell. The action is then given by
\begin{equation}
    S_{U}=\int dt \ d^{3}x \ [\dot{\Lambda}{\cal T}^{0}+\dot \pi_{i}{\cal T}^{i}+\lambda^{i}\pi_{i}+{\cal T}^{i}(\partial_{i}\Lambda)]
\end{equation}
with the Lagrange multiplier density $\lambda^i$ enforcing 3 primary constraints:
\begin{equation}\label{const2}
    \pi_i\approx 0 . 
\end{equation}
Since:
\begin{equation}\label{const3}
    \dot \pi_i=\{\pi_i, H\} =\partial_i \Lambda
\end{equation}
we now get (\ref{const1}) as a secondary constraint. No further constraints exist, and they all commute, so they are first class. 
In this rendition we have 8 new phase space variables and 4 first class constraints, so again no new local d.o.f.s:
\begin{equation}
    D=\frac{8-2\times 4}{2}=0. 
\end{equation}
It is these first class constraints that generate the gauge transformations under which the theory is invariant. 

\subsection{Method 3}
We could also gauge fix ${\cal T}^i=0$ at the expense of introducing second class constraints, writing the action as:
\be
     S_{U}=
     \int d^{4}x \,  [-\Lambda \dot{{\cal T}}^{0}- \pi_{i}\dot{{\cal T}}^{i} - \Lambda(\partial_{i}{\cal T}^{i})+\lambda^{i}\pi_{i}+\mu_{i}{\cal T}^{i}.   
\end{equation}
The last 6 constraints are then second class because $\{\pi_i,{\cal T}^j\}=\delta_i^j$. 
The number of degrees of freedom associated with this action is then counted as:
\begin{equation}
    D=\frac{8-2\times 1-3-3}{2}=0.
\end{equation}
This is the form of the action most suitable to a path integral treatment, as we explain below. 

Hence, no new local degrees of freedom emerge, a matter due to the structure of constraints, which points to the underlying gauge symmetry of the theory.  Note that the conclusion is different for {\it global degrees of freedom}, which indeed are increased by 1, as evident if we impose homogeneity and isotropy. Whereas in standard General Relativity subjected to this reduction, de Sitter space time has no degrees of freedom, in unimodular gravity there is one degree of freedom: the freedom to specify the (constant, global) value of $\Lambda$. 

\section{The Path Integral}\label{pathint}
Path integrals containing dynamical variables which obey gauge symmetries require the gauge fixing of these variables to avoid contributions from an infinite gauge volume when evaluated. Consequently, the Method 3 action is the most relevant action for path integral calculations, and so we have to deal with second class constraints. In the unimodular case, the action given by Method 3 has three primary second class constraints $\mu_{i}$ which enforce ${\cal T}^{i}\approx 0$ and three primary second class constraints $\lambda^{i}$ which enforce $\pi_{i}\approx 0$. The path integral starts off as:
\begin{align}
\begin{split}
Z=    &\int \mathcal{D}\Lambda \mathcal{D}T^{0}\mathcal{D}\lambda^{i}\mathcal{D}\mu_{i}\prod_{t}|\rm{det}\{ \xi_{a}, \xi_{b}\}|^{\frac{1}{2}}\\
    &\text{{exp}}\Biggl[\int d^{4}x (- \Lambda \dot{T}^{0}- \pi_{i}\dot{T}^{i}-\Lambda(\partial_{i}T^{i})+\lambda^{i}\pi_{i}+\mu_{i}T^{i})\Biggl],\nn
\end{split}
\end{align}
where $\rm{det}\{ \xi_{a}, \xi_{b}\}$ is the determinant of the antisymmetric matrix whose entries are the Poisson brackets of second class constraints $\xi_{a}$ and $\xi_{b}$. It was previously shown in \cite{HenneauxSlav} that this is the path integral equivalent of Dirac bracket canonical quantisation for theories containing second class constraints. In this case, as the constraints are all primary second class, the antisymmetric matrix takes the form:
\begin{equation}
    \{\xi_{a}, \xi_{b}\}= 
    \begin{vmatrix}
    \{T^{i},T^{i} \} & \{T^{i}, \pi_{j} \}\\
    \{ \pi_{j}, T^{i} \} & \{ \pi_{j}, \pi_{j} \}\\
    \end{vmatrix}= 
    \begin{vmatrix}
        0_{(3 \times 3)} &  \delta^{i}_{j(3 \times 3)}\\
        -\delta^{i}_{j(3 \times 3)} & 0_{(3 \times 3)}\\
    \end{vmatrix}\nn
\end{equation}
The determinant of this antisymmetric matrix is therefore $1$. As the constraints enforce the vanishing of the spatial components ${\cal T}^i$ and of their conjugate momenta on-shell, the path integral for the unimodular extension reduces to
\begin{equation}
Z=    \int \mathcal{D}\Lambda \mathcal{D}T^{0}\ \text{{exp}}\Biggl[i\left(\int d^{4}x \ \dot \Lambda {T}^{0}-\Bigl[\Lambda{\cal T}^{0}\Bigl]^{f}_{i}\right)\Biggl].
\end{equation}
The boundary term resulting from integration of the first term by parts has been included since it is non vanishing in general. This is exactly the minisuperspace path integral used to derive the Hartle-Hawking and Vilenkin metric representation and Chern-Simons connection representation wavefunctions of the universe~\cite{PathInt}. This is unsurprising since ${\cal T}^{i}\approx 0$ is also a consequence of enforcing homogeneity and isotropy. We stress that the results of this section are independent of the representation chosen for the base. \\

These results also hold if we choose ${\bf \alpha}=M_{P}^{2}$ as this constant has mass dimensions $[M_{P}^{2}]=M^{2}$. Choosing the base action as $(6)$, the total action is then given by:
\begin{equation}
    S=S_{0}-\int d^{4}x \ M_{P}^{2}\cdot \partial_{\mu}T^{\mu}_{R},
\end{equation}
where $T_{R}$ is the Ricci time dual to the Planck mass squared. The Ricci time variable will also have mass dimension $[T_{R}]=M^{1}$. In this case, the path integral for the unimodular extension takes the exact same form as $(24)$, with $\Lambda$ replaced by $M_{P}^{2}$. In the case where both $\Lambda$ and $M_{P}^{2}$ are both promoted to variables, the full path integral is given by:
\begin{align}
\begin{split}
    {\cal Z}=&\int {\cal D}g_{\mu \nu} \int{\cal D}\Lambda \int{\cal D}T^{0}\int{\cal D}M_{P}^{2}\int{\cal D}T^{0}_{R}\\
    &{\rm exp}\bigg[i\bigg(S_{0}+\int d^{4}x \ \dot{\Lambda}T^{0}+\int d^{4}x \ M_{P}^{2}T^{0}_{R}\\ 
    &-\bigg[\Lambda T^{0} \bigg]^{f}_{i}-\bigg[M_{P}^{2}T^{0}_{R} \bigg]^{f}_{i}\bigg)\bigg].
\end{split}
\end{align}
Both $\Lambda$ and $M_{P}^{2}$ are functions of coordinate time only. Evaluating the $T^{0}$ and $T^{0}_{R}$ path integrals leads to the delta functions $\delta(\dot{\Lambda})$ and $\delta(\dot{M_{P}})$. In general, for a function $f(t)$, the functional and normal integration measures are related by $df(t)={\cal D}f\delta(\dot{f})$. Applying this relation to the above path integral leads to the reduced path integral:
\begin{align}
\begin{split}
    {\cal Z}=\int {\cal D}g_{\mu \nu} \int d\Lambda \int dM_{p}^{2} \ {\rm exp}&\bigg[i\bigg(S_{0}-\bigg[\Lambda T^{0} \bigg]^{f}_{i}\\
    &-\bigg[M_{P}^{2}T^{0}_{R} \bigg]^{f}_{i}\bigg)\bigg].
\end{split}
\end{align}
If the initial and final values of $\Lambda$, $M_{p}^{2}$ and their dual times are chosen to be equal, then the boundary terms in $(27)$ vanish. Physically, this may correspond to an expansion of the universe from these initial values until it reaches a maximum size, then re-collapse back to the same initial state. This scenario is an important aspect of phenomenology in the Sequester model. Therefore, it is no surprise that the above path integral corresponds to the sequester path integral conjectured in \cite{padilla-review}.

\section{Proca term in unimodular-like theories}\label{Proca-Solutions}
Given the $U(1)$ symmetry of unimodular-like theories one may expect a parallel with electromagnetism, the EM potential $A^\mu$ paralleled by ${\cal T}^\mu$. This is explored in~\cite{unimodEM}, where a free kinetic is added on to unimodular theory. Here we investigate symmetry breaking in straight unimodular theory, using Proca's theory as a parallel. Recall that the Proca action in electromagnetism is defined by the addition to the usual EM action of a mass term quadratic in $A_{\mu}$:
\begin{equation}
    S_{PROCA}=\int d^{4}x \bigg(-\frac{1}{4}F_{\mu \nu}F^{\mu \nu}-\frac{M^{2}}{2}A_{\mu}A^{\mu} \bigg)
\end{equation}
(using the common assumption of giving positive energy to the spatial modes in $(A^i)^2$).  The action is then no longer invariant under the local $U(1)$ gauge symmetry defined by the transformation $A_{\mu} \rightarrow A_{\mu}+\partial_{\mu}\xi$.

Similarly, a Proca mass term for ${\cal T^{\mu}}$ can be defined. However, it must be emphasized that ${\cal T}^{\mu}$ is a vector density, so the most literal Proca term would not be a scalar (this could lead to an interesting model for local Lorentz symmetry breaking). As explained after (\ref{unimodVec}), to make ${\cal T}^\mu $ a proper
vector we must appeal to the metric, defining:
\begin{align}
    \begin{split}
        T^{\mu}&={\cal T}^{\mu}/\sqrt{-g}\\
        T_{\mu}&=g_{\mu \nu}T^{\nu}.
    \end{split}
\end{align}
A suitable scalar Proca term can then be added to unimodular theory:
\begin{eqnarray}\label{unimodProca1}
    S_{U}&=&\int d^{4}x \ \bigg(-\Lambda (\partial_{\mu}{\cal T}^{\mu})+ \frac{M^{2}}{2}\sqrt{-g}T_{\mu}T^{\mu}\bigg)\nn\\
    &=&\int d^{4}x \ \bigg(-\Lambda (\partial_{\mu}{\cal T}^{\mu})+\frac{M^{2}}{2}\frac{g_{\mu \nu}}{\sqrt{-g}}{\cal T}^{\mu}{\cal T}^{\nu}\bigg)
\end{eqnarray}
where we use the opposite convention for the sign of the mass term as that used in Proca theory, for reasons that will be obvious presently (we want to give positive energy to the scalar mode).
The added mass term respects diffeomorphism invariance while breaking the $U(1)$ gauge symmetry associated with the unimodular extension. 
We can also write the action completely in terms of $T^\mu$ as:
\begin{eqnarray}\label{unimodProca2}
    S_U&=& \int d^{4}x\sqrt{-g} \ \bigg(-\Lambda\nabla_{\mu}T^{\mu}+\frac{M^{2}}{2} T_{\mu}T^{\mu} \bigg)\nn\\
    &=& \int d^{4}x\sqrt{-g} \ \bigg((\partial_\mu\Lambda)T^{\mu}+\frac{M^{2}}{2} T_{\mu}T^{\mu} \bigg).
\end{eqnarray}
(where we have ignored a boundary term going between the two). 
Due to notational and algebraic simplicity for further calculations, the above action will be used as the unimodular Proca action. We stress that all subsequent results can be obtained from either rendition of the Proca action. 

This procedure could be applied to any other unimodular-like theory based on different $\bm\alpha$ (c.f. Eq.~\ref{genunimod}), but the dimensions of the Proca coupling would have to be adjusted.  
For a target ``constant'' $\alpha$ with mass dimensions:
\begin{equation}
    [\alpha]=M^n
\end{equation}
we have (recalling that we set $\hbar=c=1$):
\be
[T^\mu]=[{\cal T}^\mu]=M^{3-n}
\ee
and so a (gauge-invariant) zero-mode time $[T_\alpha]=[V][{\cal T}^0]=M^{-n}$ (cf. Eq.~\ref{zeroT}). 
Hence the Proca coupling has dimensions $M^{2n-2}$. In the case we have illustrated, $[\Lambda]=M^2$, so the Proca term is indeed a mass term and the $M$ used above is indeed a ``mass''. But if we  target the Planck mass squared, $\alpha=M_P^2$, so that its dual is the Ricci time used in sequesteration~\cite{pad,pad1}, then the Proca coupling is dimensionless. Also it matters which function of the constant or constants we took for $\alpha$ (that is, canonical transformations matter). For example,
if we took $\rho_\Lambda$ instead of $\Lambda$ (as is done in~\cite{pad,pad1}), then indeed the conjugate is four-volume time ($[T]=M^{-4}=L^4$), unlike the time conjugate to $\Lambda$ (which would be mixed with Ricci time in sequester scenarios). But then the coupling would have dimensions $M^6$. 

Finally, in the context of unimodular-like theories with several $\alpha_I$ and $T^\mu _I$ (where $I$ indexes the different $\bm \alpha$ in (\ref{genunimod})), we could define a general mass matrix:   
\begin{equation}
    S_U=\int d^{4}x\sqrt{-g} \ \bigg(-\alpha_I \nabla_{\mu}T_I^{\mu}+\frac{M^{2}_{IJ}}{2} T_{I \mu}T_J^{\mu} \bigg)
\end{equation}
(Einstein notation implied for the indices $I,J$). 
The fact that this matrix does not need to be diagonal, but can be diagonalized, then points to the existence of a rotation between a ``flavour'' space of constants and their canonically conjugated times, and the mass eigenmodes.  In a future publication we will explore how this can be applied to the sequester mechanics to shed light on the value of the observed cosmological constant. 

\section{Solutions and the propagating modes}\label{probs}

Breaking the symmetry releases a new propagating degree of freedom, but whereas in Proca's case this is in addition to 2 existing ones (a longitudinal polarization mode in addition to the 2 transverse ones), here the original theory has no local degrees of freedom. 

This can be illustrated by evaluating the equations of motion of (\ref{unimodProca1}) (but similar results would be obtained starting from (\ref{unimodProca2})), to find:
\begin{eqnarray}
    \nabla_{\mu}T^{\mu}&=&-M_P^2\label{EOM1}\\
    \partial_\mu\Lambda&=&  - M^2  T_\mu \label{EOM2}
\end{eqnarray}
that is, the equation of motion for $T^\mu$ remains the same (cf.~\ref{UEOM1})
but the equation for $\Lambda$ is modified (cf.~\ref{UEOM2}), with local variations in $\Lambda$ permitted. These equations can be combined into:
\begin{equation}
    \Box \Lambda= M^{2} M_P^2
\end{equation}
which is a sourced wave equation. Since this equation is linear its general solutions can be written as:
\begin{equation}
    \Lambda=\Lambda_0 + \chi
\end{equation}
with:
\begin{eqnarray}
     \Box\chi&=&0 \\
    \Box\Lambda_{\rm 0}&=&M^{2} M_P^2\label{wave2}
\end{eqnarray}
We can choose $\Lambda_0$ to be homogeneous on the surface $\Sigma$ defining the gauge-invariant zero mode that provides the time $T$. Hence $\Lambda_0$ is the (time-varying) zero mode in $\Lambda$ in this model (we will evaluate it in Section~\ref{discussion}). The mode $\chi$ is a massless scalar field, a Lambdon, as it were, upon quantization. This is the new local degree of freedom that has been released by breaking gauge invariance. 

From the point of view of the Hamiltonian analysis performed in Section~\ref{HamHT} it is not surprising that we acquire a local degree of freedom. By breaking the Gauge symmetry we lose constraints (or change their nature), increasing the number od degrees of freedom. Using Section~\ref{method2}, for example, we get the secondary constraint:
\begin{equation}\label{const3}
    S_i =\dot \pi_i=\{\pi_i, H\} =\partial_i \Lambda + M^2T_i\approx 0
\end{equation}
wherein before we had the spatial constancy of $\Lambda$. These 3 secondary constraints do not commute with the primary constraints (since $\{\pi_i,S_j\}= M^2\delta_{ij}$) so they are second class. This results in one  local degree of freedom ($(8-6)/2=1$).

Note that unlike in the formulation of~\cite{Bufalo} the use of the metric and of the vector $T^\mu$ do result in a contribution to the  Stress-Energy-Momentum tensor in Unimodular Proca theory:
\bea\label{set1}
    T_{\mu \nu}&=&\frac{-2}{\sqrt{-g}}\frac{\delta \mathcal{S}_M}{\delta g^{\mu \nu}}\nn\\
    &=& -[(\partial_\mu\Lambda)T_\nu+(\partial_\nu\Lambda)T_\mu ]+g_{\mu\nu}(\partial_\alpha\Lambda)T^\alpha +
    \nn\\
    &&- M^2\left(T_\mu T_\nu -\frac{1}{2}g_{\mu\nu}T_\alpha T^\alpha\right).
\eea
For plain unimodular theory ($M^2=0$) this is zero on-shell (since $\partial_\mu\Lambda=0$ is an equation of motion). This is not the case if $M^2\neq 0$. Eliminating the $T^\mu$ by means of (\ref{EOM2}) we get:
\begin{equation}\label{set2}
    T_{\mu \nu}=\frac{1}{M^{2}} \bigg(
   (\partial_\mu\Lambda)( \partial_\nu\Lambda)-\frac{1}{2}g_{\mu\nu}(\partial_\alpha\Lambda)( \partial^\alpha \Lambda) \bigg)
\end{equation}
This is just the stress energy tensor of a massless scalar field $\Lambda/M$. Notice that it was important to choose the sign of $M^2$ for Unimodular-Proca theory as we did, that is the opposite of the sign chosen for Electromagnetic-Proca theory. This is because our new mode is a scalar instead of a longitudinal polarization. Therefore we want to give positive energy to the scalar mode $T^0$, rather than the spatial modes $T^i$.

\section{BRST Quantisation of The original Unimodular Theory}
In Sec.~\ref{symmetry} we discussed how the gauge symmetry of unimodular-like theories is a non-standard 3-dimensional representation of the $U(1)$ gauge group. This differs from regular electromagnetism which is based on a 1-dimensional representation of the $U(1)$ gauge symmetry. However, this shared $U(1)$ gauge symmetry implies that the Becchi-Rouet-Stora-Tyutin (BRST) quantisation of these theories should be similar. To gauge-fix the Maxwell action, one deals with terms of the form $B(\partial_{\mu}A^{\mu})$, or, $B(\partial_{i}A^{i})$ where $B$ is a Nakanishi-Lautrup auxiliary field. This auxiliary field leads to a Dirac-delta function enforcing the Lorenz gauge $\partial_{\mu}A^{\mu}\approx 0$, or, the Coulomb gauge $\partial_{i}A^{i}\approx0$ on-shell. One can equivalently consider off-shell formulations of these gauge fixing conditions. In these formulations, a gauge-parameter is introduced such that the gauge fixing action is quadratic in this new parameter and the particular gauge to enforced. Substituting the equations of motion for the gauge parameter back into the action yields the original gauge fixing term.

In Sec.V, we showed that the minisuperspace path integral is the result of enforcing the gauge fixing condition $T^{i}\approx0$. Taking inspiration from the Maxwell case, in this section we will develop an off-shell formulation of this gauge fixing condition. This off-shell gauge fixing action contains a spatial Proca term which is perfectly consistent with previous results for unimodular theories. Additionally, we develop a consistent BRST quantisation of the original and off-shell gauge fixing actions. This procedure closely mirrors that of the Maxwell case.

\subsection{The Off-Shell gauge fixing action \& Spatial Proca}
From the Hamiltonian analysis we see that the on-shell vanishing of the $T^{i}$ components and the survival of $T^{0}$ component allows for the notion of a 4-volume time on-shell. The addition of a quadratic term $(T^{0})^{2}$ ruins the 4-volume time relation and leads to a contribution to the energy-momentum tensor. However, a suitable Proca term for the spatial components can be defined which respects the original unimodular relations. The off-shell formulation of the $T^{i}$ gauge fixing condition is defined as:

\begin{equation}
    S^{\rm OFF-SHELL}_{\rm GF}=-\int d^{4}x\sqrt{-g} \ \Big(\frac{1}{2\xi}T_{i}T^{i}+\frac{1}{2}\gamma_{i}\gamma^{i}\xi \Big),
\end{equation}
 where $\xi$ is a gauge parameter and $\gamma_{i}$ is an auxiliary 3-vector. Varying $S_{\rm GF}$ with respect to $\xi$ yields
 \begin{equation}
     \sqrt{\frac{T_{i}T^{i}}{\gamma_{i}\gamma^{i}}}=\xi.
 \end{equation}
As the action can be considered as quadratic in $\xi$, the above equation of motion for $\xi$ can be substituted into the action to yield:
\begin{equation}
    S_{\rm U}+ S_{\rm GF}= \int d^{4}x \ \Lambda \dot{T}^{0}+ \Lambda(\partial_{i}T^{i})-\gamma^{i}T_{i},
\end{equation}
which is the original gauge fixing condition. While not central to this analysis, a similar off-shell gauge fixing action for the conjugate momenta $\pi_{i}$ can be defined.

\subsection{The BRST Action \& BRST Symmetry}
As the path integral has been gauge fixed, the next step in BRST quantisation is to define a Fadeev-Popov ghost action. The quantisation procedure is then considered complete when the original unimodular, gauge-fixing and ghost actions all obey a defined nilpotent BRST symmetry. The BRST symmetry is determined by the  Slavnov derivative $\delta_{B}$ acting on each term in the BRST action. For the action containing the regular gauge-fixing term, this BRST symmetry is defined as
\begin{eqnarray}
      \delta_{B}\Lambda&=&0,\label{BRST1}\\ \delta_{B}T^{0}&=&0,\\
        \delta_{B}T^{i}&=&\nabla^{i}c\\
        \delta_{B}c&=&0,\\ \delta^{\alpha}_{B}\nabla^{i}\Bar{c}&=&(\gamma^{i}+\nabla^{i}\Lambda)\label{BRSTcbar}\\
        \delta_{B}\gamma^{i}&=&0.\label{BRSTlast}
\end{eqnarray}
This BRST symmetry is internally consistent as $\delta_{B}^{2}$ vanishes for all terms. The BRST action associated with the regular gauge fixing condition is given by:
\begin{align}
\begin{split}
    S_{\rm BRST}\equiv S_{\rm U}+&S_{\rm GF}+S_{\rm GH}=\\
    &\int d^{4}x\sqrt{-g}(\Lambda\nabla_{\mu}T^{\mu}-\gamma_{i}T^{i}+\nabla^{i}c\nabla_{i}\bar{c}),\nn
\end{split}
\end{align}
such that $\delta_{B}(S_{\rm BRST})=0$. There exists a number of interesting similarities and differences between the gauge fixing and ghost actions of unimodular gravity and those of electromagnetism. When gauge fixing the Maxwell action, one considers terms such as $B(\partial_{\mu}A^{\mu})$ or $B(\partial_{i}A^{i})$, as we said. These are gauge fixing conditions on the derivatives of the components of the gauge field $A^{\mu}$. 
In the unimodular case, there are three constraints $\gamma_{i}$ which directly enforce $T^{i}\approx0$. As the Slavnov derivative acts non-trivially on the spatial components only, the corresponding ghost kinetic term is purely spatial. This differs from usual electromagnetism in which the ghost kinetic term consists of the full four derivative. Additionally, to have a consistent BRST symmetry, the Slavnov derivative acting $\Bar{c}$ must be accompanied with a gradient term $\nabla^{i}$. The definition by $(51)$ can be viewed in two ways. The first is that the Slavnov derivative $\delta_{B}$ is acting on $\nabla^{i}\Bar{c}$. The second viewpoint is that $\delta_{B}\nabla^{i}$ is a unique Slavnov derivative acting on $\Bar{c}$.

A consequence of the nilpotency of the Slavnov derivative is that we can always consider a gauge transformation on the original action defined by $S_{BRST} \rightarrow S_{BRST}+\delta_{B}G_{1}$ where the Slavnov derivative $\delta_{B}$, and the integral $G_{1}$ are Grassmann odd such that $\delta_{B}G_{1}$ is Grassmann even. The overall Grassmann odd quantity $\delta_{B}(S_{BRST}+\delta_{B}G_{1})$ then vanishes. This property is most evident if we take the BRST action which contains the off-shell gauge fixing condition shown in $(44)$. In this case, the BRST action containing the spatial Proca term is given by:
\begin{equation}
    S_{BRST}^{\rm OFF-SHELL}=\int d^{4}x\sqrt{-g} \ \bigg(\Lambda \partial_{\mu}T^{\mu}-\frac{1}{2\xi}T^{i}T_{i}+\nabla_{i}\Bar{c}\nabla^{i}c \bigg).\nn
\end{equation}
The previously defined BRST symmetry (cf. Eqns.~(\ref{BRST1})-(\ref{BRSTlast})) is mostly applicable, with the only difference being 
in (\ref{BRSTcbar}), where we should define
\begin{equation}
\delta^{\beta}_{B}\nabla^{i}\Bar{c}=\bigg(\frac{1}{\xi}T^{i}+\nabla^{i}\Lambda \bigg),
\end{equation}
so that $\delta_{B}S_{BRST}=0$ for the off-shell action.
It must be noted that if we take the perspective that the $\delta_{B}$ acts on $\nabla^{i}\Bar{c}$, then the BRST symmetry is not nilpotent since $(\delta_{B}^{\beta})^{2}\nabla^{i}c\neq0$. If we adopt the second perspective that $\delta_{B}\nabla^{i}$ is an operator acting on $\Bar{c}$ then we see that:
\begin{align}
\begin{split}
    (\delta^{\beta}_{B}\nabla)^{2}\Bar{c}= &\delta_{B}^{\beta}\nabla_{i}(\delta^{\beta}_{B}\nabla^{i}\Bar{c}),\\
    =&\delta_{B}^{\beta}\nabla_{i}\bigg(\frac{1}{\xi}T^{i}+\nabla^{i}\Lambda \bigg),\\
    =&\nabla_{i}\bigg(\frac{1}{\xi}\delta_{B}^{\beta}(T^{i})+\nabla^{i}(\delta_{B}^{\beta}(\Lambda))\bigg),\\
    =&\frac{\nabla_{i}\nabla^{i}c}{\xi}=0.
\end{split}
\end{align}
Therefore the operator $\delta_{B}^{\beta}\nabla_{i}$ is only nilpotent on-shell. This is in exact analogy with the EM case. Once again, the only difference is due to the BRST symmetry only being defined for $T^{i}$. In the BRST quantisation of the Maxwell action, the quantity $\Box c/\xi$ vanishes on-shell. As the off-shell BRST action has been defined using the first term in the off-shell gauge fixing action in $(44)$, we must now define a suitable suitable integral $G_{1}$ such that $\delta_{B}G_{1}$ yields the second term in $(44)$. In this case, we will consider two integrals $G_{1}$ and $G_{2}$ defined as:
\begin{align}
\begin{split}
    G_{1}&=\int d^{4}x\sqrt{-g} \ \frac{1}{2}(-\nabla_{i}\Bar{c})\gamma^{i}\xi\\
    G_{2}&=\int d^{4}x\sqrt{-g} \ \frac{1}{2}(\nabla_{i}\Lambda)\gamma^{i}\xi
\end{split}
\end{align}
From these definitions we see that $G_{1}$ is Grassmann odd while $G_{2}$ is Grassmann even. To retrieve the second term in $(44)$, we take the Slavnov derivative of $G_{1}$ while $G_{2}$ serves purely as a counter-term to remove an unwanted part of $\delta_{B}G_{1}$. This yields:
\begin{align}
\begin{split}
    \delta_{B}G_{1}+G_{2}=&\int d^{4}x\sqrt{-g} \ \frac{1}{2}(-\delta^{\alpha}_{B}\nabla_{i}\Bar{c}+\nabla_{i}\Lambda)\gamma^{i}\xi,\\
    =&\int d^{4}x\sqrt{-g} \ \frac{1}{2}(-\gamma_{i}-\nabla_{i}\Lambda+\nabla_{i}\Lambda)\gamma^{i}\xi,\\
    =&\int d^{4}x\sqrt{-g} \ \frac{1}{2}(-\gamma_{i}\gamma^{i}\xi),
\end{split}
\end{align}
which is the second term in the off-shell gauge fixing action $(44)$. The definition of a Grassmann even counter-term, while unorthodox, is consistent with the overall idea of gauge transformations. Initially, we said that we can consider a gauge transformation $S_{BRST}\rightarrow S_{BRST}+\delta_{B}G_{1}$ such that the Grassmann odd quantity $\delta_{B}(S_{BRST}+\delta_{B}G_{1})=0$. The only other term that we could possibly add to the gauge transformation is a Grassmann even counter-term. In this case, the result is still a Grassmann even quantity $S_{BRST}+\delta_{B}G_{1}+G_{2}$. The Slavnov derivative acting on this yields the Grassmann odd quantity $\delta_{B}(S_{BRST}+\delta_{B}G_{1}+G_{2})$ which vanishes. The vanishing of $\delta_{B}G_{2}$ is ensured by $(50)$. Another difference between the BRST quantisation of the unimodular and Maxwell actions is the BRST charge of both theories. In the unimodular case, the defined BRST symmetries  of both the on-shell and off-shell actions do not contain Slavnov derivatives of $T^{0}$ or $\Lambda$. Therefore any BRST current derived from either action must only contain spatial components $J_{i}^{BRST}$. Assuming that the definition of the BRST charge is given by
\begin{equation}
    Q_{BRST}=\int d^{3}x\sqrt{-h} \ J_{0}^{BRST},
\end{equation}
then $Q_{BRST}$ is trivially zero in this case.

\section{Discussion}\label{discussion}
We close with some discussion on the phenomenological implications of these theories. This might seem
premature with such blatantly ``toy'' models, but nonetheless we will try, bearing in mind that complicating the theory (for example replacing our mass term by more general potentials) could significantly affect our statements. 
The two central predictions of Unimodular Proca theory are:
\begin{itemize}
    \item A time-variation in the zero-mode of Lambda.
    \item A massless propagating mode, or a ``Lambdon'' particle. 
\end{itemize}
Equivalent statements apply if the procedure targets any other ``constant'', such as the Planck mass or the gravitational coupling, but obviously the phenomenology is different. 

Regarding the first implication, and taking $\Lambda$ as an example, identifying $\Sigma$ with the cosmological frame, equation (\ref{wave2}) becomes:
\begin{equation*}
    \frac{1}{a^3}(a^3 \dot \Lambda_0)^.=-M^2M_P^2
\end{equation*}
where $a$ is the expansion factor and dots are derivatives with respect to proper cosmological time. Assuming the stress energy tensor of $\Lambda$ (including its Unimodular-Proca contribution (\ref{set2})) is subdominant with respect to a dominant fluid with constant equation of state $w$, we then find:
\begin{equation}\label{sol1}
    \Lambda_0=\bar \Lambda_{0}-\frac{M^2M_P^2}{2}\frac{1+w}{3+w}t^2
\end{equation}
where $\bar \Lambda_{0}$ is a constant. Hence the zero-mode of $\Lambda$ subjected to a Proca term will {\it decrease} in time, if we assume that its propagating mode is not a ghost (see discussion after (\ref{set2}); the sign of $M^2$ has this implication). Similar results apply to any other ``constant'' $\alpha$ subject to (\ref{genunimod}). The various constraints on time variations of the constants therefore translate into upper bounds on their respective $M^2$. For example, constraints from Big Bang Nucleosynthesis will translate into constraints on the mass term of a theory targeting $\alpha=M_P^2$ (since they would imply a time variation in the gravitational ``constant''). 

In the specific case of $\Lambda$ we note that its equation of state is modified, since we can read off from (\ref{set2}):
\begin{eqnarray}
    \rho_\Lambda&=&M_P^2\Lambda+\frac{\dot\Lambda^2}{2M^2}\\
    p_\Lambda&=&-M_P^2\Lambda+\frac{\dot\Lambda^2}{2M^2},
\end{eqnarray}
(in the case of $\Lambda$, but not in general, one must add to the stress energy tensor due to the Proca term, that due to  $S_0$). 
The constraints on the equation of state of dark energy can then also be translated into constraints on $M$, since it forces $w_\Lambda>-1$. The new terms could even lead to kination instead of inflation. For $\Lambda$ there is also the  possibility that it could become the dominant contribution to the cosmological stress energy tensor, in which case the solution (\ref{sol1}) would have to be modified. 

Regarding the second implication itemised above, we have to contend with the scalar particles predicted by these theories. The fact that they have not been directly seen does not mean that they have not already been ruled out by their implications. This may be the case if $\alpha$ is the gravitational coupling (or the Planck mass, if the two are identified), for the same reason that gravitational scalars can be ruled out by the milisecond pulsar, for example. However, such arguments depend crucially on the choice of $\alpha$: if this is $\Lambda$ then the Lambdon is sufficiently elusive to bypass most indirect observational constraints. 

In closing we note that there are significant differences between Electromagnetic Proca theory and Unimodular Proca theory. For example, the representation of the gauge group is very different (it has different dimension), so one cannot lift the Stueckelberg procedure used in the first case to understand the second one. Also, before the symmetry breaking terms are included, unimodular-like theories have no propagating degrees of freedom. The new propagating degree of freedom is not a longitudinal mode, to be added on to the usual two transverse modes; instead it is a new scalar mode. We are in different territory, and so the phenomenology of standard Proca theory is not applicable here.






\section{Acknowledgments}
We thank Claudia de Rham, Arkady Tseytlin and Toby Wiseman for discussions related to this paper. This work was supported by a Bell-Burnell scholarship (RI) and the STFC Consolidated Grants ST/T000791/1 and ST/X00575/1 (JM).

\end{document}